\documentclass[12pt,draftcls,onecolumn]{IEEEtran}

\IEEEoverridecommandlockouts                              

\usepackage{amsfonts,amssymb,amsmath}
\usepackage[]{graphicx,graphics}
\usepackage{tikz}
\usetikzlibrary{automata}
\usepackage{color}
\usepackage{epstopdf}

\newtheorem{definition}{Definition}

\newtheorem{corollary}{Corollary}
\newtheorem{proposition}{Proposition}
\newtheorem{theorem}{Theorem}

\begin{document}

\title{What Information Really Matters in Supervisor Reduction?}

\author{Rong Su\thanks{Rong Su is affiliated with School of Electrical and Electronic Engineering, Nanyang Technological University, 50 Nanyang Avenue, Singapore 639798. Emails: rsu@ntu.edu.sg. The support from Singapore Ministry of Education Tier 1 Academic Research Grant M4011221.040 RG84/13 is gratefully acknowledged.}}

\maketitle

\begin{abstract}
To make a supervisor comprehensible to a layman has been a long-lasting goal in the supervisory control community. One strategy is to reduce the size of a supervisor to generate a control equivalent version, whose size is hopefully much smaller than the original one so that a user or control designer can easily check whether a designed controller fulfils its objectives and requirements. After the first journal paper on this topic appeared in 1986 by Vaz and Wonham, which relied on the concept of control covers, in 2004 Su and Wonham proposed to use control congruences to ensure computational viability. This work is later adopted in the supervisor localization theory, which aims for a control equivalent distributed implementation of a given centralized supervisor. But after so many publications, some fundamental questions, which should have been addressed in the first place, have not been answered yet, namely what information is critical to ensure control equivalence, what information is responsible for size reduction, and whether the partial observation really makes things different. In this paper we will address these fundamental questions by showing that there does exist a unified supervisor reduction theory, which is applicable to all feasible supervisors regardless of whether they are under full observation or partial observation. Our theory provides a partial order over all control equivalent feasible supervisors based on their enabling, disabling and marking information, which can be used to categorize the corresponding reduction rates. Based on this result we can see that, given two control equivalent feasible supervisors, the one under full observation can always result in a reduced supervisor no bigger than that induced by a supervisor under partial observation.            
\end{abstract}


\section{Introduction\label{GD}}
In supervisory control theory (SCT) \cite{RW87} \cite{WR87}, the control problem associated with a discrete-event system (DES) is to enforce controllable and nonblocking behavior of the plant that is admissible by the specification. When applying SCT to a real application, there are two big questions that require a user to answer, that is, are we doing the right thing, and are we doing things in the right way. The first question is about the correctness of the plant and requirement models. The second question is about correctness of supervisor synthesis, which, if the computational complexity is not a concern, has been properly answered in the SCT community. When computational complexity is indeed a big concern, several efficient synthesis approaches have been proposed in the literature, e.g., \cite{FW06} \cite{SSR10} \cite{SSR12}, which can ensure correct behaviours of the closed-loop system with low computational complexity. The first question, on the other hand, has been a long-standing hurdle for the SCT to be adopted by the industry because so far there is no efficient way to identify potential errors in plant models or requirement models. The current practice is to synthesize a supervisor based on a given plant model and requirements. An empty supervisor is usually an indication that something is wrong either in the model or in the requirements, which will prompt a system designer to undertake model or requirement updates. The current SCT and its relevant tools can assist the designer to quickly locate the problems in the model that lead to emptiness of the supervisor. The real challenge is how to determine whether the plant model and the requirements are correct, when the supervisor synthesis returns a non-empty supervisor. In this case it usually requires not only syntactic correctness but also semantic correctness, i.e., the designer has to understand the true meaning and impact of every transition in the synthesized supervisor. Thus, to make a supervisor small enough for a designer to understand its function becomes extremely important.\\              

A supervisor carries two kinds of information: the key information at each state for event enabling/disabling and marking, and the information that tracks the evolution of the plant. The latter may contain some redundancy because the plant itself also carries such evolution information. In principle, it is possible to remove redundant transitional information from the supervisor, which will not mess up with the first kind of information, i.e., a reduced supervisor can still ensure the same control capability as that of the original supervisor. This is the key idea used in Vaz and Wonham's paper on supervisor reduction \cite{VW86}, which relies on the concept of control cover.  They  proved  two  useful  reduction  theorems,  and  proposed  a corresponding (exponential time) reduction algorithm. To overcome the computational complexity involved in supervisor reduction, Su and Wonham made a significant extension in \cite{SW04} by first relaxing the concept of control cover, then providing a polynomial-time reduction algorithm based on a special type of control cover called control congruence, and finally showing that the minimal supervisor problem (MSP) of computing a supervisor with minimal state size is NP-hard. A polynomial-time lower bound estimation algorithm provided in \cite{SW04} has indicated that in many applications minimum supervisors can be achieved by using control congruence in polynomial time. Since then, this reduction algorithm has been used in many case studies, and the outcomes are promising. One major application of supervisor reduction is in supervisor localization \cite{CW10}, which aims to create a control-equivalent distributed implementation of a given centralized supervisor. \\   

The supervisor reduction theory proposed in \cite{SW04} has two major setup assumptions: (1) only full observation is considered; (2) a supervisor under consideration is a sublanguage of the plant, and there exists a one-to-one mapping from the state set of the supervisor to the state set of the plant, which can be easily satisfied by applying supremal synthesis. Since then, many questions have been raised by users. For example, can we apply supervisor reduction on partially reduced supervisors, which may not necessarily be sublanguages of a given plant, or can we apply supervisor reduction in cases with partial observation? Some result has been reported in the literature about the second question, see e.g., [ZCW16]. The main objective of supervisor reduction is to ensure control equivalence between the original supervisor and a reduced supervisor. The fundamental questions are (1) \textbf{Q1:} what information ensures control equivalence, even under partial observation, and (2) \textbf{Q2:} what information determines the reduction rate, which is the main performance index of supervisor reduction. After so many years since \cite{VW86} \cite{SW04} were published, these questions are still open. In this paper we would like to provide an answer. We will first propose a generalized supervisor reduction theory, which is applicable to all feasible supervisors, regardless of whether they are under full observation or partial observation - in the latter case, a supervisor is in general not a sublanguage of the plant. We will show that for each feasible supervisor $S$ of a plant $G$, there exists a feasible supervisor $\textbf{SUPER}$ derivable from subset construction on the synchronous product of $G$ and $S$ such that all feasible supervisors that are control equivalent to $S$ with respect to $G$ and normal with respect to $S$, i.e., all transitions in those supervisors are not redundant, can be derived via quotient construction based on a properly chosen control cover on $\textbf{SUPER}$. This result will answer our first question: (\textbf{Q1}) what information is critical for ensuring control equivalence. After that, we will define a partial order ``$\preceq$'' on those feasible supervisors by using the key information about event enabling/disabling and state marking such that for any two control equivalent supervisors $S_1$ and $S_2$ with respect to $G$, if $S_1$ is  finer than $S_2$, i.e., $S_1\preceq S_2$, then the minimum reduced supervisor induced by a minimum control cover on $S_1$ is no bigger than the one induced by a minimum control cover on $S_2$. This result provides an answer to the second question: (\textbf{Q2}) what information determines the reduction rate. As a direct consequence of this result,  as long as control equivalence holds, a feasible supervisor under full observation always results in a reduced supervisor no bigger than the one induced from a supervisor under partial observation. The whole theory is independent of a specific choice of the observability definition such as observability \cite{LW88}, normality \cite{LW88} or relative observability \cite{CZW13} - these definitions are lumped into the property of control feasibility, which states that a feasible supervisor must apply the same control law to all transitional sequences which cannot be distinguished based on observations.\\

The remaining of the paper is organized as follows. In Section II, we provide preliminaries on supervisor reduction. In Section III we discuss critical information for ensuring control equivalence. Then in Section IV we talk about information that determines reduction efficiency. We draw conclusions in Section V. 

\section{Preliminaries on supervisor reduction}
Given an arbitrary finite alphabet $\Sigma$, let $\Sigma^*$ be the free monoid with the empty string $\epsilon$ being the unit element and the string concatenation being the monoid operation. Given two strings $s,t\in\Sigma^*$, $s$
is called a \emph{prefix substring} of $t$, written as $s\leq t$, if
there exists $s'\in\Sigma^*$ such that $ss'=t$, where $ss'$ denotes
the concatenation of $s$ and $s'$. Any subset $L\subseteq\Sigma^*$ is called a \emph{language}.  The \emph{prefix closure} of
$L$ is defined as $\overline{L}=\{s\in\Sigma^*|(\exists t\in L)\,
s\leq t\}\subseteq\Sigma^*$. Given
two languages $L,L'\subseteq\Sigma^*$, let
$LL':=\{ss'\in\Sigma^*|s\in L\,\wedge\, s'\in L'\}$ denote the
concatenation of two sets.  Let $\Sigma'\subseteq\Sigma$. A mapping $P:\Sigma^*\rightarrow\Sigma'^*$ is called the \emph{natural projection} with respect to $(\Sigma,\Sigma')$,
if
\begin{enumerate}
\item $P(\epsilon)=\epsilon$, 
\item $(\forall \sigma\in\Sigma)\, P(\sigma):=\left\{\begin{array}{ll} \sigma & \textrm{ if $\sigma\in\Sigma'$,}\\
    \epsilon & \textrm{ otherwise,}\end{array}\right.$ \item $(\forall s\sigma\in\Sigma^*)\, P(s\sigma)=P(s)P(\sigma)$.
\end{enumerate}
Given a language $L\subseteq\Sigma^*$, $P(L):=\{P(s)\in\Sigma'^*|s\in L\}$. The inverse image mapping of $P$ is
\[P^{-1}:2^{\Sigma'^*}\rightarrow 2^{\Sigma^*}:L\mapsto
P^{-1}(L):=\{s\in\Sigma^*|P(s)\in L\}.\] Given $L_1\subseteq\Sigma_1^*$ and $L_2\subseteq\Sigma_2^*$, the \emph{synchronous product} of $L_1$ and $L_2$
is defined as $L_1||L_2:=P_1^{-1}(L_1)\cap P_2^{-1}(L_2)$,
where $P_1:(\Sigma_1\cup\Sigma_2)^*\rightarrow\Sigma_1^*$ and $P_2:(\Sigma_1\cup\Sigma_2)^*\rightarrow\Sigma_2^*$ are natural projections. Clearly,
$||$ is commutative and associative.\\

A plant is modelled as a \emph{deterministic finite-state automaton}, $G=(X,\Sigma,\xi,x_0,X_m)$, where $X$ stands for the state set, $\Sigma$ for the
alphabet, $\xi:X\times\Sigma\rightarrow X$ for the (partial) transition function, $x_0$ for the initial state and $X_m\subseteq X$ for the
marker state set. Here we follow the notation system in \cite{Won07} and use $\xi(x,\sigma)!$ to denote that the transition $\xi(x,\sigma)$ is defined. The domain of $\xi$ can be extended to $X\times\Sigma^*$, where $\xi(x,\epsilon)=x$ for all $x\in X$, and $\xi(x,s\sigma):=\xi(\xi(x,s),\sigma)$. The \emph{closed} behavior of $G$ is defined as $L(G):=\{s\in\Sigma^*|\xi(x_0,s)!\}$, and the \emph{marked} behavior of $G$ is $L_m(G):=\{s\in L(G)|\xi(x_0,s)\in X_m\}$. $G$ is \emph{nonblocking} if $\overline{L_m(G)}=L(G)$. We say $G$ is \emph{reachable} if for each $x\in X$ there exists $s\in L(G)$ such that $\xi(x_0,s)=x$. From now one we will only consider reachable automata. We will use $|X|$ to denote the size of the state set $X$. In some circumstances, when the state set is not explicitly mentioned, we also use $|G|$ to denote the size of an automaton, which is equal to the size of its state set. \\ 

Given two finite-state automata $G_i=(X_i,\Sigma_i,\xi_i,x_{i,0},X_{i,m})$ ($i=1,2$), the \emph{synchronous product} of $G_1$ and $G_2$, denoted as $G_1||G_2$, is a (reachable) finite-state automaton \[G=(X:=X_1\times X_2,\Sigma:=\Sigma_1\cup\Sigma_2,\xi:=\xi_1\times \xi_2,x_0:=(x_{1,0},x_{2,0}),X_m:=X_{1,m}\times X_{2,m}),\]where the (partial) transition map $\xi$ is defined as follows:
\[(\forall x=(x_1,x_2)\in X)(\forall \sigma\in\Sigma)\,\xi(x,\sigma):=\left\{\begin{array}{ll} (\xi_1(x_1,\sigma),x_2) &\textrm{$\sigma\in\Sigma_1\setminus\Sigma_2$,}\\
(x_1,\xi_2(x_2,\sigma)) &\textrm{$\sigma\in\Sigma_2\setminus\Sigma_1$,}\\
(\xi_1(x_1,\sigma),\xi_2(x_2,\sigma)) & \textrm{$\sigma\in\Sigma_1\cap\Sigma_2$.}\end{array}\right.\]
It has been shown that the automaton synchronous product is commutative and associative. Thus, it can be applied to an arbitrarily finite number of finite-state automata. In this paper we will only focus finite-state automata, whose alphabets are the same. In this case, a transition is allowed in the synchronous product if all component automata allow it. \\

Let
$\Sigma=\Sigma_c\dot{\cup}\Sigma_{uc}=\Sigma_o\dot{\cup}\Sigma_{uo}$, where disjoint
$\Sigma_c$ ($\Sigma_o$) and $\Sigma_{uc}$ ($\Sigma_{uo}$) denote respectively
the sets of \emph{controllable} (\emph{observable}) and
\emph{uncontrollable} (\emph{unobservable}) events. Let $\Gamma:=\{\gamma\subseteq\Sigma|\Sigma_{uc}\subseteq\gamma\}$ be the collection of all \emph{control patterns}. A \emph{(feasible) supervisor of $G$ under partial observation $P_o:\Sigma^*\rightarrow\Sigma_o^*$} is defined as a finite-state automaton $S=(Z,\Sigma,\delta, z_o,Z_m)$ such that 
\begin{itemize}
\item $\textrm{[Control Existence] }(\forall z\in Z)\, \{\sigma\in\Sigma|\delta(z,\sigma)!\}\in\Gamma$,
\item $\textrm{[Control Feasibility] }(\forall s,s'\in L(S))\, P_o(s)=P_o(s')\Rightarrow \delta(z_0,s)=\delta(z_0,s')$.
\end{itemize} 
The first property says that a supervisor can only disable controllable events, thus, all uncontrollable events must be allowed in the control pattern (or command) at each state $z$. This property can be ensured by enforcing controllability \cite{RW87} on the closed-loop system behaviors. The second property says that a supervisor will issue the same control pattern (or command) to strings, which are observation equivalent under $P_o$. This property ensures implementation feasibility of the supervisor, and can be enforced by various types of observability properties proposed in the Ramadge-Wonham supervisor control paradigm, e.g., observability \cite{LW88}, normality \cite{LW88}, and relative observability \cite{CZW13}. It can be checked that the second property implies that
\[(\forall z\in Z)(\forall \sigma\in\Sigma_{uo})\,\delta(z,\sigma)!\Rightarrow \delta(z,\sigma)=z,\]
namely unobservable events can only be selflooped at some states, and any transition between two different states must be observable. The closed-loop behavior of the system is denoted by two languages: the closed behavior $L(G||S)=L(G)\cap L(S)$ and the marked behavior $L_m(G||S)=L_m(G)\cap L_m(S)$.  \\

\begin{figure}[htb]
    \begin{center}
      \includegraphics[width=0.40\textwidth]{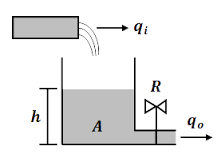}
    \end{center}
    \caption{Example 1: A single-tank system}
    \label{fig:Cyber-Security-1}
\end{figure}

\begin{figure}[htb]
    \begin{center}
      \includegraphics[width=0.90\textwidth]{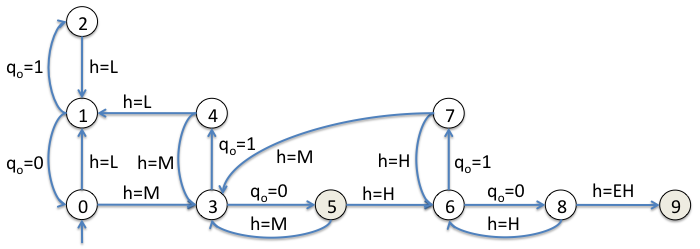}
    \end{center}
    \caption{Example 1: Automaton model of the plant $G$}
    \label{fig:Cyber-Security-2}
\end{figure}

\begin{figure}[htb]
    \begin{center}
      \includegraphics[width=0.90\textwidth]{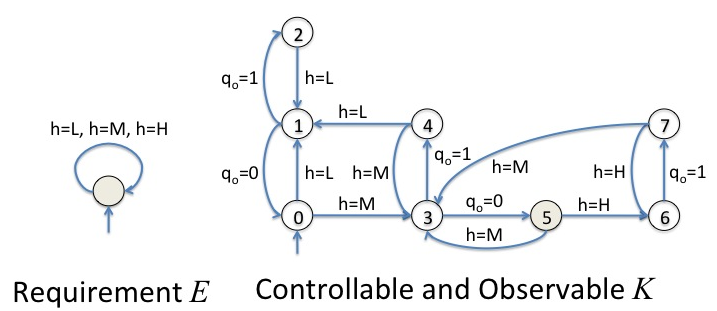}
    \end{center}
    \caption{Example 1: Automaton models of a requirement $E$ (Left) and the controllable and observable sublanguage $K$ (Right)}
    \label{fig:Cyber-Security-3}
\end{figure}

\begin{figure}[htb]
    \begin{center}
      \includegraphics[width=0.80\textwidth]{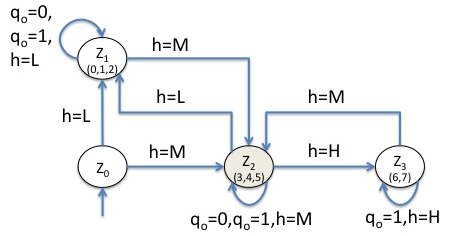}
    \end{center}
    \caption{Example 1: A feasible supervisor $S$}
    \label{fig:Supervisor-Reduction-1}
\end{figure}

To illustrate the aforementioned concepts and facilitate subsequent development, we use a simple running example of a single-tank system depicted in Figure \ref{fig:Cyber-Security-1}, which consists of  
one water supply source whose supply rate is $q_i$, one tank, and one control valve at the bottom of the tank controlling the outgoing flow rate $q_o$, whose value depends on the valve opening and the water level $h$. We assume that the valve can only be fully open or fully closed to simplify our illustration, and in case of a full opening, the water level $h$ can only go down. The water level $h$ can be measured, whose value can trigger some predefined events, denoting the water levels: \emph{low} (h=L), \emph{medium} (h=M), \emph{high} (h=H), and \emph{extremely high} (h=EH).  A simple plant model $G$ of the system is depicted in Figure \ref{fig:Cyber-Security-2},    
where the alphabet $\Sigma$ contains all events shown in the figure. The actions of opening the valve ($q_o=1$) and closing the valve ($q_o=0$) are controllable but unobservable, and all water level events are observable but uncontrollable. In the model we use a shaded oval to denote a marker state, i.e., state 5 and state 9 in Figure \ref{fig:Cyber-Security-2}. Assume that we do not want the water level to be extremely high, i.e., the event h=EH should not occur.  To prevent state 9 from being reached, we compose a requirement $E$ shown in Figure \ref{fig:Cyber-Security-3}, whose alphabet is $\{$h=L, h=M, h=H, h=EH$\}$, but the event h=EM is never allowed in the model. A controllable and observable sublanguage, i.e., a closed-loop behavior $K=L_m(G||S)$, can be synthesized by using the standard Ramadge-Wonham supervisory control paradigm, which is also depicted in Figure \ref{fig:Cyber-Security-3}.   
The corresonding feasible supervisor $S$ via subset construction on $K$ is depicted in Figure \ref{fig:Supervisor-Reduction-1}.
We can see that in $S$ all unobservable events are selflooped at some states, and transitions between different states are all labeled by observable events.\\

For a plant $G$, there may exist more than one supervisor that can fulfil a control objective, e.g., to ensure the closed-loop system behavior to be contained in a predefined requirement language $E\subseteq\Sigma^*$. Two supervisors $S_1$ and $S_2$ of $G$ are \emph{control equivalent} \cite{SW04} if $L(G||S_1)=L(G||S_2)$ and $L_m(G||S_1)=L_m(G||S_2)$. Let $\mathcal{C}(G,S)$ be the collection of all feasible supervisors of $G$ under partial observation $P_o$, which are control equivalent to a given supervisor $S$. It is desirbale to find one supervisor $S_*\in\mathcal{F}(G,S)$ such that for all $S'\in\mathcal{F}(G,S)$ we have $|S_*|\leq |S'|$, i.e., the supervisor $S_*$ has the minimum number of states. Unfortunately, it has been shown in \cite{SW04} that finding $S_*$ is NP-hard, even for a supervisor under full observation, which relies on the concept of \emph{control covers} - each control cover is a group of states in $S$ that are ``control compatible'',whose exact meaning will be explained shortly. Thus, by groupong those compatible states of $S$ together, we may get a new supervisor $S'$ such that (1) $S'$ is control equivalent to $S$; (2) $|S'|< |S|$ (ideally, $|S'|\ll |S|$). In the next couple of sections we will investigate which information is responsible for control consistency, and which is for size reduction.    

\section{Information that ensures control equivalence} 
Given a plant $G=(X,\Sigma,\xi,x_0,X_m)$ and a supervisor $S=(Z,\Sigma,\delta,z_0,Z_m)$, at each state $z\in Z$ there are four pieces of information shown below:
\begin{itemize}
\item Let $En:Z\rightarrow 2^{\Sigma}$ with 
\[z\mapsto En(z):=\{\sigma\in\Sigma|\delta(z,\sigma)!\}\]
be the ($S$-)enabled event set at state $z\in Z$. 
\item Let $D:Z\rightarrow 2^{\Sigma}$ with
\[z\mapsto D(z):=\{\sigma\in\Sigma|\neg\delta(z,\sigma)!\wedge (\exists s\sigma\in L(G))\, \delta(z_0,s)=z\}\]
be the ($S$-)disabled event set at state $z\in Z$. 
\item Let $M:Z\rightarrow\{true,false\}$ with 
\[z\mapsto M(z):=true\textrm{ if }(\exists s\in L_m(G||S))\, \delta(z_0,s)=z\]
be the $S$-marking indicator at state $z\in Z$. 
\item Let $T:Z\rightarrow\{true,false\}$ with
\[z\mapsto T(z):=true\textrm{ if }(\exists s\in L_m(G))\, \delta(z_0,s)=z\] 
be the $G$-marking indicator at state $z\in Z$.
\end{itemize}
The $(S-)$enabled event sets can be easily obtained by simply checking the transition structure of $S$. To determine other sets for each state $z\in Z$, we can first construct the synchronous product $G||S$, and then check each state tuple $(x,z)$ in the product associated with the state $z\in Z$.\\ 

As an illustration, let's revisit that supervisor $S$ for the single-tank system depicted in Figure \ref{fig:Supervisor-Reduction-1}. By undertaking the synchronous product $G||S$ we can obtain the transition structure recognizing $K$ shown in the right picture of Figure \ref{fig:Cyber-Security-3}. From that structure we can get the following:
\begin{itemize}
\item $En(z_0)=\{$h=L, h=M$\}$, $D(z_0)=\varnothing$, $M(z_0)=false$, $T(z_0)=false$, 
\item $En(z_1)=\{$q$_0$=0, q$_0$=1, h=L, h=M$\}$, $D(z_1)=\varnothing$, $M(z_1)=false$, $T(z_1)=false$, 
\item $En(z_2)=\{$q$_0$=0, q$_0$=1, h=L, h=M, h=H$\}$, $D(z_2)=\varnothing$, $M(z_2)=true$, $T(z_2)=true$, 
\item $En(z_3)=\{$q$_0$=1, h=M, h=H$\}$, $D(z_3)=\{$q$_0$=0$\}$, $M(z_3)=false$, $T(z_3)=false$.\\ \end{itemize}

Let $\mathcal{R}\subseteq Z\times Z$ be a binary relation, where $(z,z')\in\mathcal{R}$ iff the following two properties hold:
\begin{enumerate}
\item $En(z)\cap D(z')=En(z')\cap D(z)=\varnothing$,
\item $T(z)=T(z')\Rightarrow M(z)=M(z')$.
\end{enumerate}
The first condition requires that any event enabled at one state cannot be disabled at the other state. The second condition requires that both states must have the same marking status, if they are reachable by strings from the marked behavior of $G$. Notice that $\mathcal{R}$ is not transitive, thus, it is not an equivalence relation. Any two states satisfying $\mathcal{R}$ may potentially be merged together, if their suffix behaviors are ``compatible'', which is precisely captured in the following concept.\\

\begin{definition}\textnormal{A \emph{cover} $\mathcal{C}=\{Z_i\subseteq Z|i\in I\}$ of $Z$ is a \emph{control cover} on $S$ if
\begin{enumerate}
\item $(\forall i\in I)\, Z_i\neq\varnothing\wedge (\forall z,z'\in Z_i)\, (z,z')\in\mathcal{R}$,
\item $(\forall i\in I)(\forall \sigma\in\Sigma)(\exists j\in I)[(\forall z\in Z_i)\delta(z,\sigma)!\Rightarrow \delta(z,\sigma)\in Z_j]$.\hfill $\Box$\\
\end{enumerate}}
\end{definition}

Given a control cover $\mathcal{C}=\{Z_i\subseteq Z|i\in I\}$ on $S$, we construct an induced supervisor $S_{\mathcal{C}}=(I,\Sigma,\kappa,i_0,I_m)$, where $i_0\in I$ such that $z_0\in Z_{i_0}$, $I_m:=\{i\in I|Z_i\cap Z_m\neq\varnothing\}$, and $\kappa:I\times\Sigma\rightarrow I$ is the partial transition map such that for each $i\in I$ and $\sigma\in\Sigma$, $\kappa(i,\sigma):=j$ if $j$ is chosen to satisfy the following property:
\[(\exists z\in Z_i)\delta(x,\sigma)\in Z_j\wedge [(\forall z'\in Z_i)\, \delta(z',\sigma)!\Rightarrow \delta(z',\sigma)\in Z_j].\]
In general, there may exist more than one choice of $j$ satisfying the above property. A random selection among multiple choices is usually adopted. We now have the first result.\\

\begin{theorem}\textnormal{$S_{\mathcal{C}}$ constructed above is a feasible supervisor, which is control equivalent to $S$. \hfill $\Box$}\end{theorem}
Proof: 1.	We first claim that $L_m(G||S)\subseteq L_m(G||S_{\mathcal{C}})$.
Let  $s\in L_m(G||S)$.  If  $s = \epsilon$,  then  $z_0 \in Z_m$.  Since  $z_0\in Z_{i_0}$,  we  have $Z_{i_0}  \cap Z_m \neq  \varnothing$. Therefore $i_0 \in I_m$, namely $\epsilon\in L_m(S_{\mathcal{C}})$. Let $s = \sigma_0\cdots \sigma_k$ ($k > 0$). Because

\[\delta(z_0, \sigma_.0)!,\, \delta(z_0, \sigma_0\sigma_1)!, \cdots , \delta(z_0, \sigma_0\sigma_1\cdots \sigma_k)!,\]
we have

\[\delta(z_0, \sigma_0)!\textrm{ and } \delta(z_j, \sigma_j)!\textrm{ with }z_{j+1} = \delta(z_0, \sigma_0\cdots \sigma_j),\,	j= 1,\cdots, k\]
Since $\{Z_i l i\in I\}$ is a control cover on $Z$, by Definition 1 and the definition of $\kappa$ we have
\[(\forall j:0\leq j\leq k)(\exists i_j,i_{j+1}\in I)z_j\in Z_{i_j}\wedge z_{j+1}\in Z_{i_{j+1}}\wedge \kappa(i_j,\sigma_j)=i_{j+1}.\]
Therefore, $\kappa(i_0, s)!$. Since $s\in L_m(G||S)$, we have $\kappa(i_0, s)\in Z_m$. Therefore $s\in L_m(G||S_{\mathcal{C}})$, namely 
\[L_m(G||S) \subseteq L_m(G||S_{\mathcal{C}}).\]
By taking the prefix closure on both sides, and recall that $\overline{L_m(G||S)}=L(G||S)$, we have
\[L(G||S) \subseteq L(G||S_{\mathcal{C}}).\]

\noindent 2.	For the reverse inclusion, let $s\in L(G||S_{\mathcal{C}})$. If $s = \epsilon$ then, as $L(G||S)\neq\varnothing$, $s\in L(G||S)$. Suppose $s = \sigma$. Then $\kappa(i_0, s)!$, so there are $z\in Z_{i_0}$ and $z'\in Z$ such that $\delta(z, \sigma) = z'$, namely $\sigma\in En_{S_{\mathcal{C}}}(z)$. By the definition of the control cover $\mathcal{C}$, $\sigma\notin  D_{i_0} (z_0)$, so either $\delta(z_0, \sigma)!$ or
\[(\forall t\in\Sigma^*)\delta(z_0, t) = z_0\Rightarrow t\sigma\notin L(G).\]
But since $s=\sigma\in L(G||S)$, we conclude $\delta(z_0, \sigma)!$, namely $s\in L(G||S)$. Of course, by definition of the control cover $\mathcal{C}$, there follows $\delta(z_0, \sigma) = z'\in Z_{i'}$ for some $i'\in I$. In general, let $s = \sigma_0\sigma_1\cdots \sigma_k$. Repeating the foregoing argument k-fold, we see that $s\in L(G||S_{\mathcal{C}})$ implies $s\in L(G||S)$. This shows that $L(G||S_{\mathcal{C}})\subseteq L(G||S)$.

\noindent 3. Let $s\in L_m(G||S_{\mathcal{C}})$. As shown above, $\delta(z_0, s)!$ with $\delta(z_0, s)=z \in \kappa(i_0, s)$. Since $\kappa(i_0, s)\in I_m$, there exists $z'\in Z_{\kappa(i_0 ,s)}\cap X_m$, namely $M_S(z') = true$. By the definition of control cover, we know that there  is   $s'\in L_m(G||S)$    such  that
$\delta(z_0, s') = z'$, namely $T_S(z')=true$. At the same time, $s\in L_M(G||S)$ implies $T_S(z)=true$. By definition of control cover $\mathcal{C}$, we get $M_S(z) = M_S(z') = true$,
namely $\delta(z_0, s)=z\in Z_m$, and $s\in L_m(G||S)$, as required.

So far we have shown that $L(G||S)=L(G||S_{\mathcal{C}})$ and $L_m(G||S)=L_m(G||S_{\mathcal{C}})$. Finally, we need to show that $S_{\mathcal{C}}$ is a feasible supervisor, namely those two conditions must hold. The Control Existence condition obviously hold because the construction of $S_{\mathcal{C}}$ from $S$ does not disable any event more than $S$ does. Since $S$ is feasible, namely the Control Existence condition holds, we know that this condition must hold for $S_{\mathcal{C}}$. For the second condition of Control Feasibility, notice that all unobservable events are selflooped at some states in $S$, by the definition of control cover $\mathcal{C}$, it is clear that those unobservable events are also selflooped in some states in $S_{\mathcal{C}}$. Thus, the Control Feasibility condition holds for $S_{\mathcal{C}}$, which completes the proof.\hfill $\blacksquare$\\

\begin{figure}[htb]
    \begin{center}
      \includegraphics[width=0.60\textwidth]{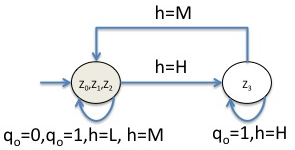}
    \end{center}
    \caption{Example 1: An induced supervisor $S_{\mathcal{C}}$}
    \label{fig:Supervisor-Reduction-2}
\end{figure}

Theorem 1 indicates that we can start with any given plant $G$ and feasible supervisor $S$ to generate another feasible supervisor $S'$, which is control equivalent to $S$ with respect to $G$, by applying the aforementioned construction induced by a properly chosen control cover on $S$. The interesting part of this story is that we do not need to know how we get that $S$ in the first place. Thus, we have a unified way of undertaking supervisor reduction regardless of whether $S$  is under full observation or partial observation. As an illustration, let's revisit that single-tank system, whose feasible supervisor $S$ is depicted in Figure \ref{fig:Supervisor-Reduction-1}. Based on the aforementioned analysis about those four sets, i.e., $En(z)$, $D(z)$, $M(z)$ and $T(z)$, for each state $z\in Z$, we can check that the set $\mathcal{C}:=\{\{z_0,z_1,z_2\},\{z_3\}\}$ is a control cover. The resulting induced supervisor $S_{\mathcal{C}}$ is depicted in Figure \ref{fig:Supervisor-Reduction-2}.
We can easily check that $S_{\mathcal{C}}$ is control equivalent to $S$ with respect to $G$. From $S_{\mathcal{C}}$ we can see that what $S$ really does is to prevent the valve from being closed when the water level is high, which matches our expectation perfectly.\\

Next, we will present a result similar to the Generalized Quotient Theorem in \cite{SW04}. \\

\begin{definition}\textnormal{Given a plant $G$ and a feasible supervisor $S$, let $S'=(Z',\delta',\Sigma,z_0',Z_m')$ be another feasible supervisor of $G$. Then $S'$ is \emph{normal} with respect to $S$ if the following hold:
\begin{enumerate}
\item $(\forall z\in Z')(\forall \sigma\in\Sigma) \delta'(z,\sigma)!\Rightarrow (\exists s\sigma\in L(G||S))\, \delta'(z_0',s)=z$,
\item $(\forall z\in Z_m')(\exists s\in L_m(G||S))\,\delta'(z_0',s)=z$.\hfill $\Box$\\
\end{enumerate}}\end{definition} 

\begin{definition}\textnormal{Given automata $G_A=(X_A,\Sigma,\xi_A,x_{A,0},X_{A,m})$ and $G_B=(X_B,\Sigma,\xi_B,x_{B,0},X_{B,m})$, $G_A$ is \emph{DES-epimorphic} to $G_B$ under DES-epimorphism $\theta:X_A\rightarrow X_B$ if
\begin{enumerate}
\item $\theta$ is surjective,
\item $\theta(x_{A,0})=x_{B,0}$ and $\theta(X_{A,m})=X_{B,m}$,
\item $(\forall x,x'\in X_A)(\forall \sigma\in\Sigma)\xi_A(x,\sigma)=x'\Rightarrow \xi_B(\theta(x),\sigma)=\theta(x')$,
\item $(\forall x\in X_B)(\forall \sigma\in\Sigma)\xi_B(x,\sigma)!\Rightarrow (\exists x'\in\theta^{-1}(x))\xi_A(x',\sigma)!$.
\end{enumerate}
In particular, $G_A$ is \emph{DES-isomorphic} to $G_B$ if $\theta:X_A\rightarrow X_B$ is bijective.\hfill $\Box$\\
}\end{definition}

Given a plant $G$ and a feasible supervisor $S$, by computing the synchronous product of $G$ and $S$, i.e., $G||S$, we can obtain the closed-loop (closed and marked) behaviours. It is well know that, by applying subset construction on $G||S$ with respect to $P_o:\Sigma^*\rightarrow \Sigma_o^*$ followed by insering selfloops of projected unobservable events at appropriate states in the resulting automton, we can derive a feasible supervisor, say $\textbf{SUPER}$, which can be shown control equivalent to $S$. The following main result shows that any normal feasible supervisor, which is control equivalent to $S$ with respect to $G$, can be constructed from $\textbf{SUPER}$ by using a proper control cover on $S$.\\

\begin{theorem}\textnormal{Let $\textbf{SUPER}$ be constructed as above. Then for any normal feasible supervisor $\textbf{SIMSUP}$ with respect to $S$, which is control equivalent to $S$ with respect to $G$, there exists a control cover $\mathcal{C}$ on $\textbf{SUPER}$ such that some induced feasible supervisor $S_{\mathcal{C}}$ is DES-isomorphic to $\textbf{SIMSUP}$.\hfill $\Box$}\end{theorem}
Proof: With $\textbf{SUPER} = (Z, \Sigma, \delta, z_0, Z_m)$ and $\textbf{SIMSUP} = (Y, \Sigma, \eta, y_0, Y_m)$, for each $y\in Y$, let
\[Z(y) :=\{z\in Z l (\exists s\in L(G||S)\delta(z_0, s) = z \wedge \eta((y_0, s) = y\}\]
and define $\mathcal{C} := \{Z(y) l y\in Y\}$. We now check that $\mathcal{C}$ is a control cover on $\textbf{SUPER}$.

By normality of $\textbf{SIMSUP}$, we have $Z(y)\neq\varnothing$ for all $y\in Y$. Since $\textbf{SUPER}$ is obtained by the subset construction, for each $z\in Z$, there is $s\in L(G||S)=L(G||\textbf{SIMSUP})$ with $\delta(z_0, s) = z$ and $\eta(y_0, s)!$. Hence, $z\in Z(\eta(y_0,s))$. This shows that $\mathcal{C}= \{Z(y)l y\in Y\}$ covers $Z$.

Next,	fix $y\in Y$	and	let $a, b\in Z(y)$	with	$\sigma\in En_{\textbf{SUPER}}(a)$. We need to show that $\sigma\notin D_{\textbf{SUPER}}(b)$. Since $\textbf{SUPER}$ is constructed via subset construction, we know that for all $s\in L(G||S)$, there exists $s'\in P_o^{-1}(P_o(s))\cap L(G||S)$ such that $s'\sigma\in L(G||S)$. In addition, $\delta(z_0,s')=a$. Since $a\in Z(y)$, there exists $\hat{s}\in L(G||S)$ such that $\delta(z_0,\hat{s})=a$ and $\eta(y_0,\hat{s})=y$. Thus, we know that there exists $\hat{s}'\in  P_o^{-1}(P_o(\hat{s}))\cap L(G||S)$ such that $\hat{s}'\sigma\in L(G||S)$ and $\delta(z_0,\hat{s}')=a$. Since $\textbf{SIMSUP}$ is a feasible supervisor, we know 
that $\eta(y_0,\hat{s}')=y$. Thus, $\eta(y,\sigma)!$. Since $b\in Z(y)$, there exists $t\in L(G||S)$ such that $\delta(z_0,t)=b$ and $\eta(y_0,t)=y$. If there exists $t'\sigma\in L(G)$ such that $\delta(z_0,t')=b$, we know that there must exist $\hat{t}\in P_o^{-1}(P_o(t))\cap L(G||S)$ such that $\hat{t}\sigma\in L(G)$, $\delta(z_0,\hat{t})=b$ and, because $\textbf{SIMSUP}$ is a feasible supervisor, we have $\eta(y_0,\hat{t})=y$. Since $\hat{t}\sigma\in L(G||\textbf{SIMSUP})=L(G||\textbf{SUPER})$, we know that $\delta(b,\sigma)!$. Thus, $\sigma\notin D_{\textbf{SUPER}}(b)$, namely 
$En_{\textbf{SUPER}}(a)\cap D_{\textbf{SUPER}}(b)=\varnothing$, as required.

Next, we show that 
\[T_{\textbf{SUPER}}(a)=T_{\textbf{SUPER}}(b)\Rightarrow M_{\textbf{SUPER}}(a)=M_{\textbf{SUPER}}(b).\]
To this end, let $y\in Y$ and $a,b\in Z(y)$ with $M_{\textbf{SUPER}}(a)\neq M_{\textbf{SUPER}}(b)$. Without loss of generality, assume that $M_{\textbf{SUPER}}(a)=true$ and $M_{\textbf{SUPER}}(b)=false$. Since $M_{\textbf{SUPER}}(a)=true$, there exists $s\in L_m(G||S)$ such that $\delta(z_0,s)=a$. Thus, $T_{\textbf{SUPER}}(a)=true$. Since $a\in Z(y)$, we know that there exists $s'\in L(G||S)$ such that $\delta(z_0,s')=a$ and $\eta(y_0,s')=y$. Due to the subset construction, we know that there exists $\hat{s}\in P_o^{-1}(P_o(t))\cap L_m(G||S)$ such that $\delta(z_0,\hat{s})=a$ and, because $\textbf{SIMSUP}$ is a feasible supervisor, we have $\eta(y_0,\hat{s})=y$. This means $y\in Y_m$. Since $b\in Z(y)$, for all $t\in L(G||S)$ with $\delta(z_0,t)=b$, due to the subset construction and $\textbf{SIMSUP}$ is a feasible supervisor, we can deduce that there exists $\hat{t}\in L(G||S)$ such that $\delta(z_0,\hat{t})=b$, $\eta(y_0,\hat{t})=y$ and $t\in L_m(G)\iff \hat{t}\in L_m(G)$. Since
$M_{\textbf{SUPER}}(b)=false$, we know that $\hat{t}\notin L_m(G||S)=L_m(G||\textbf{SIMSUP})$. Since $y\in Y_m$, we can deduce that $\hat{t}\notin L_m(G)$. Thus, $t\notin L_m(G)$. Since $t$ is arbitrarily chosen, we know that $T_{\textbf{SUPER}}(b)=false$. Thus, we have
\[M_{\textbf{SUPER}}(a)\neq M_{\textbf{SUPER}}(b)\Rightarrow T_{\textbf{SUPER}}(a)\neq T_{\textbf{SUPER}}(b),\]
which is equivalent to
\[T_{\textbf{SUPER}}(a)=T_{\textbf{SUPER}}(b)\Rightarrow M_{\textbf{SUPER}}(a)=M_{\textbf{SUPER}}(b).\]

Finally, we need to show that for each $y\in Y$ and $\sigma\in\Sigma$, there exists $y'\in Y$ such that
 \[(\forall z\in Z(y))\delta(z,\sigma)!\Rightarrow \delta(z,\sigma)\in Z(y').\]
Let $z\in Z(y)$ and $\delta(z,\sigma)!$. Clearly, there exists $s\sigma\in L(G||S)$ such that $\delta(z_0,s)=z$. By using an argument similar as above, we know that there exists $s'\in P_o^{-1}(P_o(s))\cap L(G||S)$ such that $\delta(z_0,s')=z$, $\eta(y_0,s')=y$, and $s'\sigma\in L(G||S)$. Clearly, $\eta(y,\sigma)!$. Thus, $\delta(z,\sigma)\in Z(\eta(y,\sigma))$, as required. 

So far we have shown that $\mathcal{C}$ is a control cover on $\textbf{SUPER}$. By Theorem 1 we know that an induced $S_{\mathcal{C}}$ is a feasible supervisor, which is control equivalent to $S$ with respect to $G$. In addition, there exists a natural DES-isomorphism 
\[\theta:Y\rightarrow 2^{Z}:y\mapsto \theta(y):=Z(y).\] 
Thus, $S_{\mathcal{C}}$ is DES-isomorphic to $\textbf{SIMSUP}$, which completes the proof. \hfill $\blacksquare$\\

Up to now we have developed a general theory on supervisor reduction, which unifies both the full observation case and the partial observation case. As a matter of fact, we can see that the concrete way of ensuring observability in a feasible supervisor is not important in achieving control equivalence during supervisor reduction. By knowing the plant $G$ and a feasible supervisor $S$ will be sufficient for us to construct a feasible supervisor, which is control equivalent to $S$, and hopefully has a (significantly) smaller size.   

\section{Information that determines reduction efficiency} 
Our case studies indicate that a supervisor with full observation usually allows a much higher reduction rate than what a supervisor with partial observation allows. An interesting question is what causes such discrepancy. In this section we will try to answer this question, which provides a deep insight on the actual effects of full/partial observations on supervisor reduction.\\ 

Given a plant $G$ and a feasible supervisor $S$, each feasible supervisor $S'\in\mathcal{F}(G,S)$ carried four pieces of critical information captured by $(En_{S'},D_{S'},M_{S'},T_{S'})$. We define a partial order ``$\preceq$'' among elements of $\mathcal{F}(G,S)$, where for all $S_i=(Z_i,\Sigma,\delta_i,z_{i,0},Z_{i,m})\in\mathcal{F}(G,S)$ ($i=1,2$), we say $S_1$ is \emph{finer than} $S_2$, denoted as $S_1\preceq S_2$, if for all $s\in L(G||S)$ let $z_1:=\delta_1(z_{1,0},s)$ and $z_2=\delta_2(z_{2,0},s)$, and we have 
\begin{itemize}
\item $En_{S_1}(z_1)\subseteq En_{S_2}(z_2)$ and $D_{S_1}(z_1)\subseteq D_{S_2}(z_2)$,
\item $M_{S_1}(z_1)=true\Rightarrow M_{S_2}(z_2)=true$,
\item $T_{S_1}(z_1)= true\Rightarrow T_{S_2}(z_2)=true$.
\end{itemize}
In other words, $S_1$ is finer than $S_2$ if for each pair of states $z_1$ in $S_1$ and $z_2$ in $S_2$ reachable by the same string in $L(G||S)$, the enaling and disablig event sets at $z_1$ are subsets of those at $z_2$, and the values of the $S$-marking indicator and the $G$-marking indicator at $z_1$ are true imply that those values at $z_2$ are also true. Informally speaking,
$S_1$ carries less redundant (or finer) information than what $S_2$ does, in terms of ensuring control equivalence.\\

\begin{figure}[htb]
    \begin{center}
      \includegraphics[width=0.80\textwidth]{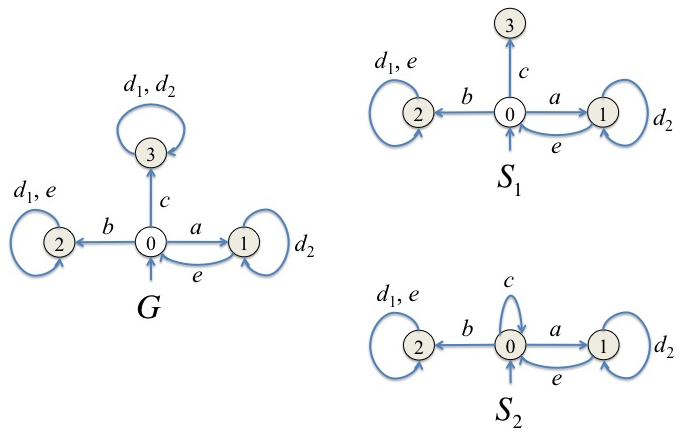}
    \end{center}
    \caption{Example 2: A plant $G$ (left), supervisors $S_1$ (right top) and $S_2$ (right bottom)}
    \label{fig:Supervisor-Reduction-3}
\end{figure}

We now use a simple example depicted in Figure \ref{fig:Supervisor-Reduction-3} to illustrate this idea of partial order over control equivalent feasible supervisors. The alphabet of the plant $G$ is
$\Sigma=\{a,b,c,d_1,d_2,e\}$, $\Sigma_c=\{d_1,d_2\}$, and all events are observable for the sake of simplicity. It is not difficult to check that $S_1$ and $S_2$ are control equivalent, which essentially disable events $d_1$ and $d_2$ after firing the event $c$. To check that $S_1$ is finer than $S_2$, we notice that we only need to check those conditions for two strings $s=\epsilon$ and $s=c$ because for other strings in $L(G||S)$, $S_1$ and $S_2$ are the same. For $s=\epsilon$, we have $z_1=0$ in $S_1$ and $z_2=0$ in $S_2$. Clearly, $En_{S_1}(z_1)=\{a,b,c\}= En_{S_1}(z_2)$, and $D_{S_1}(z_1)=\varnothing=D_{S_2}(z_2)$. In addition, we can check that $M_{S_1}(z_1)=false$ and  $M_{S_2}(z_2)=true$, and $T_{S_1}(z_1)=false$ and $T_{S_2}(z_2)=true$. Thus, those conditions hold for $s=\epsilon$. For $s=c$ we have $z_1=3$ in $S_1$ and $z_2=0$ in $S_2$. Clearly, $En_{S_1}(z_1)=\varnothing\subseteq En_{S_1}(z_2)=\{a,b,c\}$, and $D_{S_1}(z_1)=\{d_1,d_2\}=D_{S_2}(z_2)$. In addition, we can check that $M_{S_1}(z_1)=true=M_{S_2}(z_2)$, and $T_{S_1}(z_1)=true=T_{S_2}(z_2)$. Thus, we can conclude that $S_1$ is finer than $S_2$. \\

\begin{proposition}\textnormal{Given a plant $G$ and a feasible supervisor $S$, let $\textbf{SUPER}=(\hat{Z},\Sigma,\hat{\delta},\hat{z}_0,\hat{Z}_m)$ be constructed above. Then for all $s\in L(G||S)$, let $z=\hat{\delta}(\hat{z}_0,s)$ and we have the following:  
\begin{enumerate}
\item $En_{\textbf{SUPER}}(z)=\{\sigma\in\Sigma|(\exists s'\sigma\in L(G||S))\, P_o(s)=P_o(s')\}$,
\item $D_{\textbf{SUPER}}(z)=\{\sigma\in\Sigma|(\exists s'\sigma\in L(G))\, P_o(s)=P_o(s')\wedge s'\in L(S)\wedge s'\sigma\notin L(S)\}$.\hfill $\Box$ 
\end{enumerate}}\end{proposition}
Proof: Recall that $\textbf{SUPER}$ is obtained by applying subset construction on $G||S$. Thus, we know that the following properties hold:
\begin{itemize}
\item[(a)] $(\forall \sigma\in En_{\textbf{SUPER}}(z))(\exists s'\sigma\in L(G||S))\, \hat{\delta}(\hat{z}_0,s')=z$,  
\item[(b)] $(\forall s',s''\in L(G||S)) P_o(s')=P_o(s'')\Rightarrow \hat{\delta}(\hat{z}_0,s')=\hat{\delta}(\hat{z}_0,s'')$,
\item[(c)] for any two strings $s',s''\in L(G||S)$, if $\hat{\delta}(\hat{z}_0,s')=\hat{\delta}(\hat{z}_0,s'')$, then
\[\{\sigma\in\Sigma|(\exists t\sigma\in L(G||S))P_o(s')=P_o(t)\}=\{\sigma'\in\Sigma|(\exists t'\sigma\in L(G||S))P_o(s'')=P_o(t')\}.\] 
\end{itemize}
Thus, we know that
\begin{eqnarray*}
En_{\textbf{SUPER}}(z) & = & \cup_{s'\in L(G||S):\hat{\delta}(\hat{z}_0,s')=z}\{\sigma\in\Sigma|s'\sigma\in L(G||S)\} \textrm{ by Property (a)}\\
&= & \cup_{s'\in L(G||S):\hat{\delta}(\hat{z}_0,s')=z} \{\sigma\in\Sigma|(\exists t\sigma\in L(G||S))P_o(s')=P_o(t)\}\textrm{ by Property (b)}\\
& = &  \{\sigma\in\Sigma|(\exists s'\sigma\in L(G||S))P_o(s)=P_o(s')\}\textrm{ by Property (c)}
\end{eqnarray*}
from which we conclude that statement 1) is true.

To show statement 2), let $\sigma'\in D_{\textbf{SUPER}}(z)$. Then $\neg\hat{\delta}(z,\sigma')!$ but there exists $s'\sigma'\in L(G)$ such that $\hat{\delta}(\hat{z}_0,s')=z$. Clearly, $s'\in L(G||S)$ but $s'\sigma'\notin L(S)$. In addition, due to the natural of subset construction, we can choose $s'$ in such a way that $P_o(s)=P_o(s')$. Thus, we know that 
\[\sigma'\in \{\sigma\in\Sigma|(\exists s'\sigma\in L(G))\, P_o(s)=P_o(s')\wedge s'\in L(S)\wedge s'\sigma\notin L(S)\},\]
which means $D_{\textbf{SUPER}}(z)\subseteq\{\sigma\in\Sigma|(\exists s'\sigma\in L(G))\, P_o(s)=P_o(s')\wedge s'\in L(S)\wedge s'\sigma\notin L(S)\}.$
To show the opposite direction of set inclusion, let $\sigma'\in \{\sigma\in\Sigma|(\exists s'\sigma\in L(G))\, P_o(s)=P_o(s')\wedge s'\in L(S)\wedge s'\sigma\notin L(S)\}$. Then there exists $s'\sigma'\in L(G)$ such that $P_o(s)=P_o(s')$, $s'\in L(S)$ and $s'\sigma'\notin L(S)$. By Property (b) we know that $\hat{\delta}(\hat{z}_0,s')=z$. Since $S$ is a feasible supervisor, by the property of control feasibility, we know that for all $s''\in L(S)$ with $P_o(s)=P_o(s'')$, if $s''\sigma\notin L(S)$. Thus, we can conclude that $\neg\hat{\delta}(z,\sigma)!$. Thus, $\sigma'\in D_{\textbf{SUPER}}(z)$, which means
$D_{\textbf{SUPER}}(z)\supseteq\{\sigma\in\Sigma|(\exists s'\sigma\in L(G))\, P_o(s)=P_o(s')\wedge s'\in L(S)\wedge s'\sigma\notin L(S)\}.$\hfill $\blacksquare$\\

\begin{theorem}\textnormal{Given a plant $G$ and a feasible supervisor $S$, let \textbf{SUPER} be constructed above. Then for all $S'\in\mathcal{F}(G,S)$, we have $\textbf{SUPER}\preceq S'$.\hfill $\Box$}\end{theorem}
Proof: For an arbitrary feasible supervisor $S'\in\mathcal{F}(G,S)$, Properties (a)-(b) in Proposition 1 still hold. But Property (c) does not necessarily hold. For this reason, by (the first part of the proof of) Proposition 1, it is not difficult to see that for all $s\in L(G||S)$ let $z_1=\hat{\delta}(\hat{z}_0,s)$ and $z_2=\delta_2(z_{2,0},s)$, and we have $En_{\textbf{SUPER}}(z_1)\subseteq En_{S_2}(z_2).$ By using a similar argument as in the second part of the proof in Proposition 1, and the fact that the choice of $s'$ to ensure $P_o(s)=P_o(s')$ may not be feasible for an arbitrary feasible supervisor $S'$, we can easily conclude that $D_{\textbf{SUPER}}(z_1)\subseteq D_{S_2}(z_2)$. By definitions of functions of $M$ and $T$, we can check that $M_{\textbf{SUPER}}(z_1)=true$ implies $M_{S_2}(z_2)=true$, and $T_{\textbf{SUPER}}(z_1)=true$ implies $T_{S_2}(z_2)=true$. Thus, we have $\textbf{SUPER}\preceq S'$.\hfill $\blacksquare$\\

Theorem 3 indicates that for all feasible supervisors in $\mathcal{F}(G,S)$, $\textbf{SUPER}$ has the finest information, which still ensures control equivalence. The interesting point is that for any feasible supervisor $S'\in\mathcal{F}(G,S)$, we can construct $\textbf{SUPER}$ by applying subset construction on $G||S'$, namely we can always obtain the finest feasible supervisor, which is control equivalent to $S$ with respect to $G$. Nevertheless, the size of $\textbf{SUPER}$ could be big for a practical application. Thus, supervisor reduction may be directly applied to any attainable feasible supervisor $S'\in\mathcal{F}(G,S)$. The following result indicates that the supervisor reduction rate, which is defined as the ratio of the size of a (minimally) reduced supervisor and the size of the supervisor that we start with, solely depends on the fineness of the key information specified by those four functions - the finer the information, the higher the reduction rate.\\

\begin{theorem}\textnormal{Given a plant $G$ and a feasible supervisor $S$, let $S_1, S_2\in\mathcal{F}(G,S)$ be normal with respect to $S$, and assume that $S_1\preceq S_2$. Let $\mathcal{C}_1$ and $\mathcal{C}_2$ be minimum control covers of $S_1$ and $S_2$ respectively. Then $|\mathcal{C}_1|\leq |\mathcal{C}_2|$.\hfill $\Box$}\end{theorem}
Proof: Let $S_j=(Z_j,\Sigma,\delta_j,z_{j,0},Z_{j,m})$ ($j=1,2$), and $\mathcal{R}_j\subseteq Z_j\times Z_j$ the compatibility binary relation. Let $\mathcal{C}_2=\{Z_{2,i}\subseteq Z_2|i\in I_2\}$ be a minimum \emph{control cover} on $S_2$. By Definition 1,
\begin{enumerate}
\item $(\forall i\in I_2)\, Z_{2,i}\neq\varnothing\wedge (\forall z,z'\in Z_{2,i})\, (z,z')\in\mathcal{R}_2$,
\item $(\forall i\in I_2)(\forall \sigma\in\Sigma)(\exists j\in I_2)[(\forall z\in Z_{2,i})\delta_2(z,\sigma)!\Rightarrow \delta_2(z,\sigma)\in Z_{2,j}]$.
\end{enumerate}
Since $S_2$ is normal with respect to $S$, we can derive that for each $z\in Z_2$ there exists $s\in L(G||S)$ such that $\delta_2(z_{2,0},s)=z$. For each $Z_{2,i}\in\mathcal{C}_2$, let \[\mathcal{L}(Z_{2,i}):=\{s\in L(G||S)|\delta_2(z_{2,0},s)\in Z_{2,i}\}\cup (\Sigma^*\setminus L(G)).\] We can easily check that \[En_{S_2}(Z_{2,i}):=\cup_{z\in Z_{2,i}}En_{S_2}(z)=\{\sigma\in\Sigma|s\sigma\in L(G||S)\wedge s\in \mathcal{L}(Z_{2,i})\}.\]
Since $S_1,S_2\in\mathcal{F}(G,S)$, we know that $\mathcal{L}(Z_{2,i})\subseteq L(G||S_2)=L(G||S_1)$. Let
\[\hat{\mathcal{C}}_1:=\{Z_{1,i}\subseteq Z_1|[z\in Z_{1,i}\iff (\exists s\in \mathcal{L}(Z_{2,i}))\delta_1(z_{1,0},s)=z]\wedge i\in I_2\}.\] 
We now show that $\hat{\mathcal{C}}_1$ is a control cover of $S_1$. First, we show that $\hat{\mathcal{C}}_1$ is a cover of $Z_1$. To see this, notice that $\cup_{i\in I_2}\mathcal{L}(Z_{2,i})=L(G||S)=L(G||S_2)=L(G||S_1)$. Since $S_1$ is also normal with respect to $S$, we know that $\hat{\mathcal{C}}_1$ must be a cover of $Z_1$. 

To show that $\hat{\mathcal{C}}_1$ is a control cover of $S_1$, we need to show that those two conditions hold. To see the satisfaction of the first condition, for each $Z_{1,i}\in\hat{\mathcal{C}}_1$ and for all $z_1,z_1'\in Z_{1,i}$, we know that there exist $s,s'\in\mathcal{L}(Z_{2,i})$ such that $\delta_1(z_{1,0},s)=z_1$ and $\delta_1(z_{1,0},s')=z_1'$. On the other hand, let $z_2=\delta_2(z_{2,0},s)$ and $z_2'=\delta_2(z_{2,0},s')$. Since $S_1\preceq S_2$, we know that 
\begin{itemize}
\item $En_{S_1}(z_1)\subseteq En_{S_2}(z_2)$ and $D_{S_1}(z_1)\subseteq D_{S_2}(z_2)$,
\item $M_{S_1}(z_1)=true\Rightarrow M_{S_2}(z_2)=true$,
\item $T_{S_1}(z_1)=true\Rightarrow T_{S_2}(z_2)=true$,
\end{itemize} 
and 
\begin{itemize}
\item $En_{S_1}(z_1')\subseteq En_{S_2}(z_2')$ and $D_{S_1}(z_1')\subseteq D_{S_2}(z_2')$,
\item $M_{S_1}(z_1')=true\Rightarrow M_{S_2}(z_2')=true$,
\item $T_{S_1}(z_1')=true\Rightarrow T_{S_2}(z_2')=true$.
\end{itemize}
Since $(z_2,z_2')\in\mathcal{R}_2$, we have
\begin{itemize}
\item $En_{S_2}(z_2)\cap D_{S_2}(z_2')=En_{S_2}(z_2')\cap D_{S_2}(z_2)=\varnothing$,
\item $T_{S_2}(z_2)=T_{S_2}(z_2')\Rightarrow M_{S_2}(z_2)=M_{S_2}(z_2')$.
\end{itemize}
Thus, we can easily conclude that
\[En_{S_1}(z_1)\cap D_{S_1}(z_1')=En_{S_1}(z_1')\cap D_{S_1}(z_1)=\varnothing.\]
To show that 
\[T_{S_1}(z_1)=T_{S_1}(z_1')\Rightarrow M_{S_1}(z_1)=M_{S_1}(z_1'),\]
it is clear that if $T_{S_1}(z_1)=T_{S_1}(z_1')=false$, then by the definition of $M_{S_1}$ we know that $M_{S_1}(z_1)=M_{S_1}(z_1')=false$. So we only need to show that when $T_{S_1}(z_1)=T_{S_1}(z_1')=true$, we have $M_{S_1}(z_1)=M_{S_1}(z_1')$. Suppose it is not true. Then with loss of generality, let $M_{S_1}(z_1)=true$ and $M_{S_1}(z_1')=false$. Since $M_{S_1}(z_1')=false$ and $T_{S_1}(z_1')=true$, we can conclude that $M_{S_2}(z_2')=false$ due to the control equivalence of $S_1$ and $S_2$. But on the other hand, since $S_1\preceq S_2$, we know that $M_{S_1}(z_1)=true$ implies that $M_{S_2}(z_2)=true$. Thus, we have  $T_{S_2}(z_2)=T_{S_2}(z_2')=true$, $M_{S_2}(z_2)=true$, and $M_{S_2}(z_2')=false$, which contradicts our assumption that 
\[T_{S_2}(z_2)=T_{S_2}(z_2')\Rightarrow M_{S_2}(z_2)=M_{S_2}(z_2').\]
Thus, we can only have $M_{S_1}(z_1)=M_{S_1}(z_1')$,
which means $(z_1,z_1')\in\mathcal{R}_1$.

To see the satisfaction of the second condition, for each $i\in I_2$, $\sigma\in\Sigma$, we know that there exists $j\in I_2$ such that 
\[(\forall z\in Z_{2,i})\delta_2(z,\sigma)!\Rightarrow \delta_2(z,\sigma)\in Z_{2,j}.\]
For each $z'\in Z_{1,i}$, if $\delta_1(z',\sigma)!$, there there are two cases. Case 1: there exists $s\in\mathcal{L}(Z_{2,i})$ such that $\delta_1(z_{1,0},s)=z'$ and $s\sigma\in L(G||S)$. Since $\delta_2(z_{2,i},s)=z''\in Z_{2,i}$ and $\delta_2(z'',\sigma)!$, we know that $s\sigma\in \mathcal{L}(Z_{2,j})$. Thus, $\delta_1(z',\sigma)\in Z_{1,j}$. Case 2: for all $s'\in\mathcal{L}(Z_{2,i})$ with $\delta_1(z_{1,0},s')=z'$, we have  $s'\sigma\notin L(G||S)$. Then clearly $s'\sigma\notin L(G)$ because otherwise the first condition of control cover will be violated. Thus, we still have that $s'\sigma\in\mathcal{L}(Z_{2,j})$. Thus, $\delta_1(z',\sigma)\in Z_{1,j}$. So in either case, we can conclude that 
\[(\forall z\in Z_{1,i})\delta_1(z,\sigma)!\Rightarrow \delta_1(z,\sigma)\in Z_{1,j},\]
which completes our proof that $\hat{\mathcal{C}}_1$ is a contol cover of $S_1$. 

Clearly, $|\hat{\mathcal{C}}_1|=|\mathcal{C}_2|$. On the other hand, if $\mathcal{C}_1$ is a minimum control cover of $S_1$, we know that $|\mathcal{C}_1|\leq |\hat{\mathcal{C}}_1|$. Thus, we can conclude that $|\mathcal{C}_1|\leq |\mathcal{C}_2|$. \hfill $\blacksquare$\\    

As an illustration, in Example 2 depicted in Figure \ref{fig:Supervisor-Reduction-3} we know that $S_1\preceq S_2$. We can easily compute $\hat{S}_1$ and $\hat{S}_2$, which are the minimum feasibles supervisors control equivalent to $S_1$ and $S_2$ respectively.  The results are shown in Figure \ref{fig:Supervisor-Reduction-4} below.
\begin{figure}[htb]
    \begin{center}
      \includegraphics[width=0.80\textwidth]{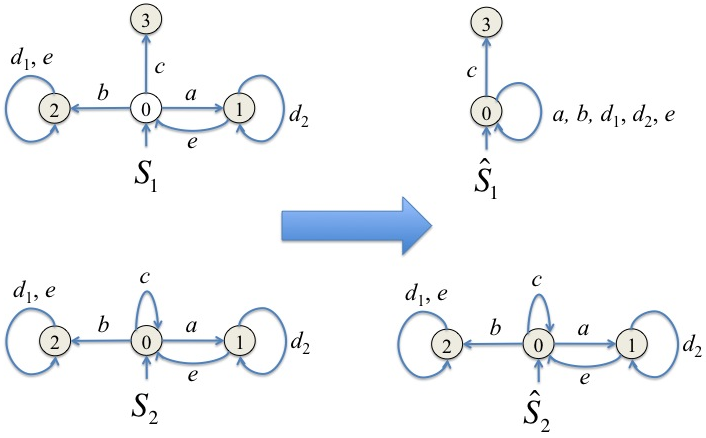}
    \end{center}
    \caption{Example 2: Reduced supervisors $\hat{S}_1$ (right top) and $\hat{S}_2$ (right bottom)}
    \label{fig:Supervisor-Reduction-4}
\end{figure}
It is clear that $|\hat{S}_1|=2< |\hat{S}_2|=3$, which matches the conclusion made in Theorem 4.\\

With Theorem 1 and Theorem 4 we are finally able to answer the  question: why the reduction rate is higher for a supervisor under full observation than that for a supervisor under partial observation. Given a plant $G$ and a feasible supervisor $S_f$, if $L_m(G||S_f)$ happens to be observable with respect to $(G,P_o)$ \cite{LW88} for some observable alphabet $\Sigma_o\subseteq\Sigma$, then there exists another feasible supervisor $S_p$ such that $S_p$ is control equivalent to $S_f$ with respect to $G$, namely $L(G||S_f)=L(G||S_p)$ and $L_m(G||S_f)=L_m(G||S_p)$. But notice that $S_f$ and $S_p$ work under different observation scenarios. The interesting part is that the same supervisor reduction procedure can be applied to both $S_f$ and $S_p$, which again indicates that a specific choice of observability to derive $S_p$ is not critical. We have the following result.\\

\begin{corollary}\textnormal{Given a plant $G$, let $S_f$ and $S_p$ be constructed above. Assume that $S_f$ is DES-isomorphic to $G||S_f$ and $S_p$ is DES-isomorphic to the subset construction of $G||S_p$. Let $\textbf{SIMSUP}_f$ and $\textbf{SIMSUP}_p$ be the minimum reduced supervisors of $S_f$ and $S_p$ respectively, based on control cover construction. Then we have $|\textbf{SIMSUP}_f|\leq |\textbf{SIMSUP}_p|$.\hfill $\Box$}\end{corollary}
Proof: Since $S_f$ is DES-isomorphic to $G||S_f$ and $S_p$ is DES-isomorphic to the subset construction of $G||S_p$, both $S_f$ and $S_p$ are the finest supervisor of their kinds. Since $S_f$ and $S_p$ are control equivalent with respect to $G$, it is not difficult to check that for all $s\in L(G||S_f)=L(G||S_p)$, let $z_f:=\delta_f(z_{f,0},s)$ and $z_p=\delta_p(z_{p,0},s)$, and we have 
\begin{enumerate}
\item $En_{S_f}(z_f)\subseteq En_{S_p}(z_p)=\cup_{s'\in P_o^{-1}(P_o(s))\cap L(G||S_p)}En_{S_f}(\delta_f(z_{f,0},s'))$,
\item $D_{S_f}(z_f)\subseteq D_{S_p}(z_p)=\cup_{s'\in P_o^{-1}(P_o(s))\cap L(G||S_p)}D_{S_f}(\delta_f(z_{f,0},s'))$,
\item $M_{S_f}(z_f)= M_{S_p}(z_p),\, T_{S_f}(z_f)= T_{S_p}(z_p)$.
\end{enumerate}
Thus, we can derive that $S_f\preceq S_p$, which by Theorem 4 we can derive that the minimum control covers $\mathcal{C}_f$ of $S_f$ and $\mathcal{C}_p$ of $S_p$ satisfie $|\mathcal{C}_f|\leq |\mathcal{C}_p|$. Thus, by Theorem 1, we know that $|\textbf{SIMSUP}_f|\leq |\textbf{SIMSUP}_p|$. \hfill $\blacksquare$\\

Corollary 1 indicates that, for two control equivalent feasible supervisors, the one under full observation always results in a (typically much) smaller reduced supervisor than what the one under partial observation can achieve. For example, in the aforementioned Example 2, no matter whether the event $c$ is observable or unobservable, the closed-loop behavior $L_m(G||S)$ is always controllable and observable, thus, $S_1$ and $S_2$ depicted in Figure \ref{fig:Supervisor-Reduction-3} can be considered as supervisors under full observation and partial observation, respectively. It is clear that the supervisor $S_1$ under full observation results in a smaller reduced supervisor $\hat{S}_1$, which is control equivalent to $S_1$.

\section{Conclusions}
So far we have developed a generalized supervisor reduction theory, which is applicable to all feasible supervisors, regardless of whether they are under full observation or partial observation. We have shown that the generalized quotient theorem in \cite{SW04} for supervisors with full observation has a counterpart in the generalized reduction theory, which states that for each feasible supervisor $S$ of a plant $G$, there exists a feasible supervisor $\textbf{SUPER}$ derivable from subset construction on $G||S$ such that all feasible supervisors that are control equivalent to $S$ with respect to $G$ and normal with respect to $S$ can be derived via quotient construction based on a properly chosen control cover on $\textbf{SUPER}$. In addition, we have provided a specific way of ordering those feasible supervisors by using the key information described in those four functions such that for any two control equivalent supervisors $S_1$ and $S_2$ with respect to $(G,S)$, if $S_1$ is finer than $S_2$, i.e., $S_1\preceq S_2$, then the minimum reduced supervisor induced from $S_1$ is no bigger than the one induced from $S_2$. As a direct consequence of this result together with Theorem 1 on the quotient construction, we know that, as long as control equivalence holds, a feasible supervisor under full observation always results in a reduced supervisor no bigger than the one induced from a supervisor under partial observation. Our theory indicates that a specific choice of observability, e.g., observability, normality or relative observability, does not play any significant role in supervisor reduction - they are all lumped into the property of control feasibility.

\end{document}